\documentclass[aps,prl,twocolumn,showpacs,superscriptaddress,amsmath,amssymb]{revtex4}

\usepackage{graphicx}
\usepackage{epstopdf}
\begin{document}

\title{Measurement of Magnetic Exchange in Ferromagnet-Superconductor Bilayers of La$_{2/3}$Ca$_{1/3}$MnO$_3$/YBa$_2$Cu$_3$O$_7$}

\author{S.~R. Giblin} 
\affiliation{ISIS Facility, Rutherford Appleton Laboratory, Chilton, Oxfordshire OX11 0QX, United Kingdom}
\author{J.~W. Taylor} 
\affiliation{ISIS Facility, Rutherford Appleton Laboratory, Chilton, Oxfordshire OX11 0QX, United Kingdom}
\author{J.~A. Duffy}
\affiliation{Department of Physics, University of Warwick, Coventry CV4
7AL, United Kingdom}
\author{M.~W. Butchers}
\affiliation{Department of Physics, University of Warwick, Coventry CV4 
7AL, United Kingdom}
\author{C. Utfeld} 
\affiliation{H.~H.~Wills Physics Laboratory, University of Bristol, Bristol BS8 1TL, United Kingdom}
\author{S.~B. Dugdale}
\affiliation{H.~H.~Wills Physics Laboratory, University of Bristol, Bristol BS8 1TL, United Kingdom}
\author{T.~Nakamura}
\affiliation{Japan Synchrotron Radiation Research Institute, SPring-8, 1-1-1 Kouto, Sayo, Hyogo 679-5198, Japan}   
\author{C.~Visani}
\altaffiliation{Current address: Unité Mixte de Physique CNRS/Thales, 1 Avenue A. Fresnel, 91767 Palaiseau Cedex, France}
\affiliation{GFMC, Departmento Fisica Aplicada $III$, Universidad de Complutense de Madrid, 28040 Madrid, Spain}
\author{J.~Santamaria}
\affiliation{GFMC, Departmento Fisica Aplicada $III$, Universidad de Complutense de Madrid, 28040 Madrid, Spain}

\date{\today}

\begin{abstract}
The existence of coherent magnetic correlations in the normal phase of cuprate high temperature superconductors has proven difficult to measure directly. Here we report on a study of ferromagnetic/superconductor bilayers of La$_{2/3}$Ca$_{1/3}$MnO$_3$/YBa$_2$Cu$_3$O$_7$ (LCMO/YBCO), with varying YBCO layer thicknesses.  Using X-ray magnetic circular dichroism it is demonstrated that the ferromagnetic layer induces a Cu magnetic moment in the adjacent high temperature superconductor.  For thin samples, this moment exists at all temperatures below the Curie temperature of the LCMO layer. However, for a YBCO layer thicker than 12 unit cells, the Cu moment is suppressed for temperatures above the superconducting transition, suggesting this to be a direct measurement of magnetic coherence in the normal state of a superconducting oxide.

\end{abstract}
  
\maketitle

The origins of high-temperature superconductivity and the rich phase diagrams in complex oxides are still a matter of contention that have stimulated many novel experimental studies and observations. Recently the improvement of layer-by-layer growth techniques of thin films has enabled investigations of both bulk and surface properties. For most common superconductors the order parameter is thought to be antagonistic to that of the exchange mechanism in ferromagnets. Accurately grown thin films have enabled these competing interactions to be probed experimentally.  In particular, the growth of epitaxial oxide layers, with well-characterized atomically flat interfaces, consisting of superconducting layers of YBa$_2$Cu$_3$O$_7$ (YBCO) and lattice-matched ferromagnetic La$_{2/3}$Ca$_{1/3}$MnO$_3$ (LCMO) has flourished \cite{zef1,zef2,pena,huber}. Measurements on YBCO and LCMO bilayers, trilayers and superlattices have allowed studies of the competition between both long range orders \cite{pena1,pena2, nemes,soltan}.  Understanding the state of matter from which high-temperature superconductivity arises could assist  in understanding its physical origin and behavior in both bulk and thin film samples. Indeed, many surprising physical properties have been observed in the normal phase of these materials, in particular the well-defined superconducting dome in the phase diagrams of oxide materials is thought to be intimately related to properties of the pseudogap phase \cite{timusk,norman}. 

Along with advances in growth methods, many developing experimental techniques have allowed the individual layer properties to be probed directly. The individual magnetic layer properties of such superlattices can be successfully investigated by X-ray magnetic circular dichroism (XMCD), an element-specific spectroscopic technique. XMCD has been used to help argue for orbital reconstruction at the YBCO/LCMO interface \cite{chak1} and to determine the existence of a Mn-O-Cu superexchange at the superlattice interface below the ferromagnetic ordering temperature ($\approx$200K) of LCMO \cite{chak2}. This observation of an induced copper moment in the YBCO layer suggests that a systematic investigation as a function of superconductor layer thickness would be worthwhile to determine whether the induced Cu moment can be modified due to local magnetic correlations within the superconducting layer, which are known to exist in the bulk system but are difficult to characterize\cite{norman}.

In this Letter we demonstrate that magnetic correlations in the normal phase of optimally doped thin films of YBCO can be suppressed by an adjacent ferromagnetic layer when the thickness of the superconductor layer is less than a value of 12 unit cells (u.c.), corresponding to a thickness of 14nm.  For thinner YBCO layers, the ferromagnetic LCMO layer is able to induce a Cu moment in the normal state, but it does not do so for layers thicker than this.  This thickness dependence of the exchange coupling above the superconducting transition temperature strongly suggests the presence of magnetic fluctuations in the Cu planes of the YBCO film which compete with the induced superexchange interaction from the LCMO layer.

Epitaxial bilayers were grown on (100)-orientated SrTiO$_3$ substrates using a high-pressure DC sputtering system in
a pure O$_2$ atmosphere. The sample interfaces have previously been shown to be atomically 
flat with negligible diffusion and little structural distortion \cite{zef1,zef2}. In all the samples studied, the LCMO thickness was 
fixed at 40 u.c. and was grown on top of the $c-$axis orientated YBCO to form the bilayer. A total of 14 samples were investigated with YBCO thicknesses of 2, 4, 5, 6, 7, 8, 11, 15, 16, 20, 22, 25, 26 and 30 u.c.. 
The samples were characterized using X-ray reflectivity 
to confirm the structural integrity and the change of thickness expected from the growth conditions, and magnetization measurements were performed using a commercial Quantum Design
MPMS XL SQUID magnetometer. Soft X-ray absorption spectroscopy (XAS)  and soft XMCD were performed  at  
beamline BL25SU at the SPring-8 synchrotron in Japan. 
All the XMCD experiments were performed using the total electron yield (TEY) method. In all experiments the magnetic field was applied at an angle of 30 degrees to the plane of the sample. All XAS measurements were performed in an applied field of 1580Oe and the hysteresis measurements performed in applied fields up to 5000Oe.

\begin{figure}
\includegraphics[width=0.9\linewidth,clip=true, viewport=0in 0in 13.8in 21.5in]{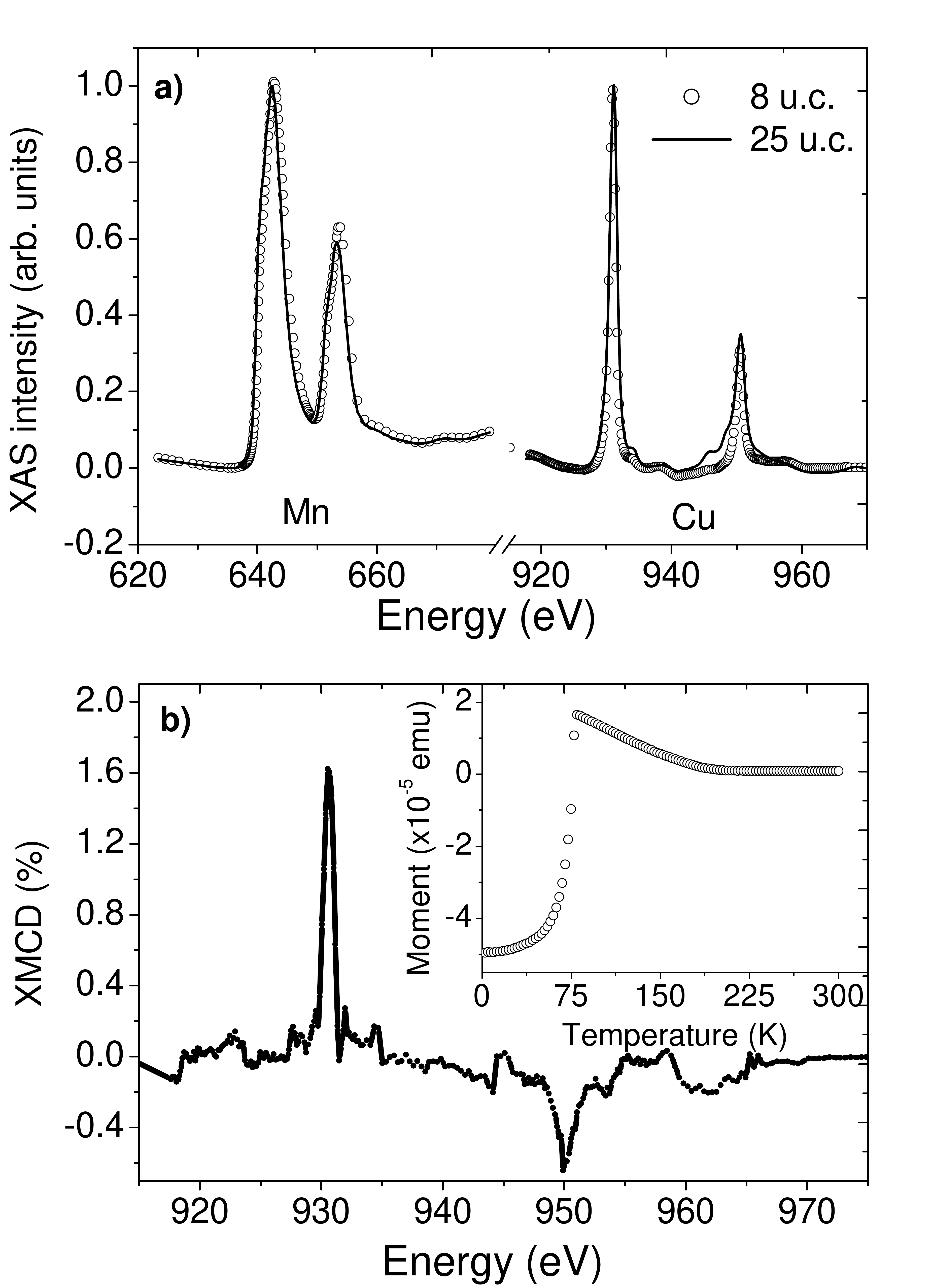}
\caption{\label{fig1} \textbf{a.} The energy dependence of the normalized XAS for two samples with a YBCO thickness
of 8 u.c. (open circles) and 25 u.c. (line) for both the Mn and Cu  L$_2$ and L$_3$  edges. The data are measured at 100K. \textbf{b.} The energy dependence of the normalized Cu edge XMCD for a sample with a YBCO thickness
of 8 u.c. measured at 100K. The L$_2$ and L$_3$ edge are clearly visible. The inset shows the temperature dependence of the magnetization for this 8 u.c. YBCO sample.
The sample was cooled and measured in 100 Oe.  The Curie temperature ($\approx{210K}$) of LCMO and the onset
of superconductivity ($\approx{80K}$) are clearly observed.}
\end{figure}

\begin{figure}
\includegraphics[width=1\linewidth,clip=true, viewport=0.0in 0.0in 13in 21in]{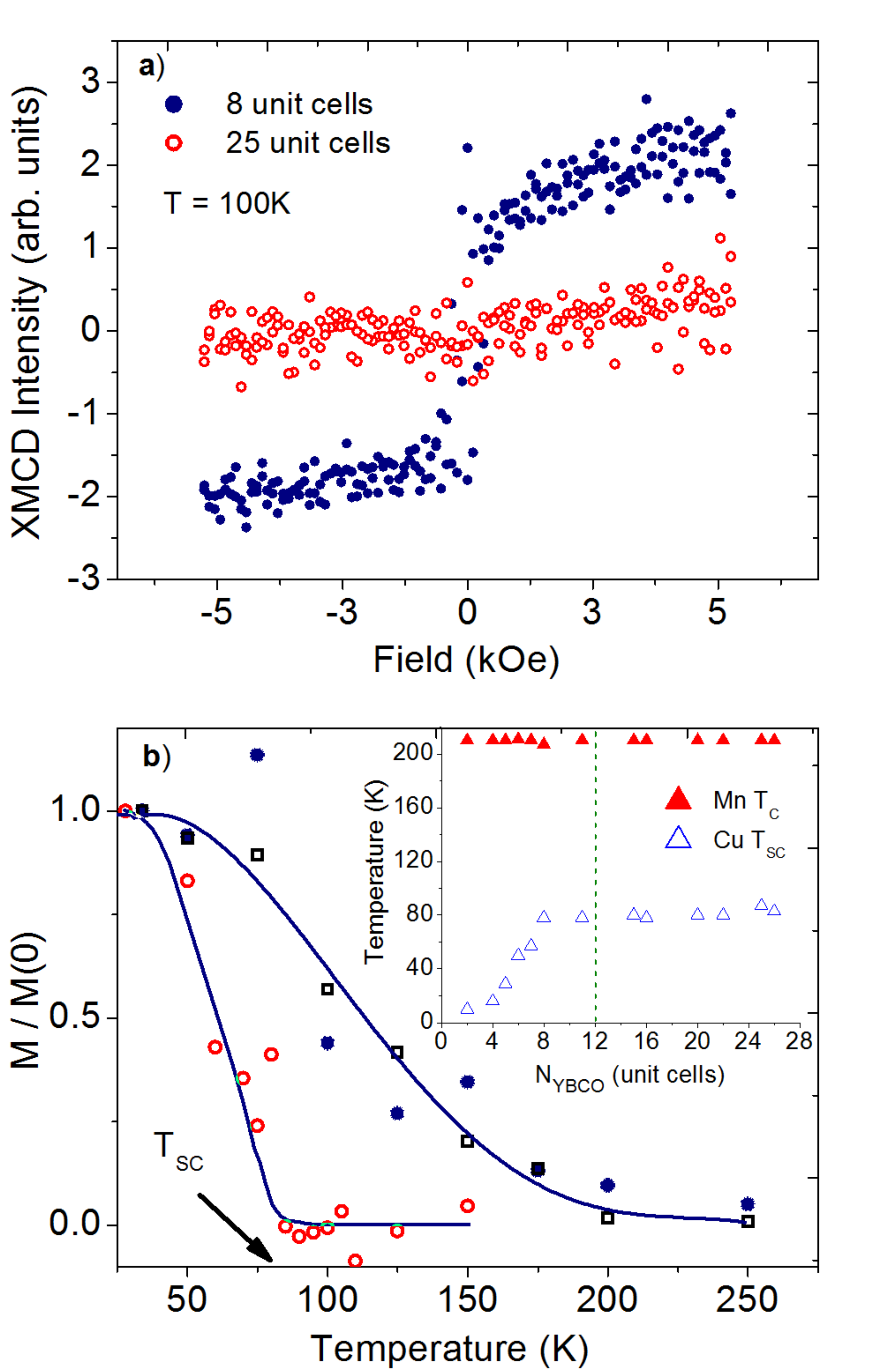}
\caption{\label{fig2}  (Color online).
\textbf{a.} The field dependent 
hysteresis for  samples with an YBCO thickness of 8 u.c. (solid blue circles) and 25 u.c. (open red circles) measured at 100K.  As 
described in the text, only samples of thickness greater than 12 unit cells of YBCO show a Cu moment above T$_{SC}$.  \textbf{b.} 
Temperature dependence of the XMCD signals for samples with a YBCO thickness of
8 u.c and 25 u.c. For 8 u.c, both the Cu (solid blue circles) and Mn (open squares) moment for the normalized spectra 
are shown. For the 25 u.c. sample only the Cu (open red circles) moment is shown. The arrow represents the superconducting transition, and the lines are guides to the eye.
The inset shows the Mn ordering at T${_C}$ and T$_{SC}$
for each sample obtained from magnetization measurements. The dotted green line splits the Cu magnetic properties of the samples as described in the text.}
\end{figure}

Fig. \ref{fig1}a shows the energy-dependent XAS for two representative samples with YBCO thicknesses of 8 u.c. and 25 u.c., at 
a temperature of 100K.  The Mn XAS is identical in both cases and the peak is consistent with the correct valence state of Mn in LCMO \cite{val}.
The Cu XAS is observed for both samples through 40 u.c. of LCMO and in both 
cases the L$_{2,3}$ edges at 951eV and 931eV are expected for CuO \cite{bg}.  Given the well-characterized interfaces, the Cu signal is very unlikely to be a consequence of Cu diffusion into the 
LCMO layers \cite{zef1,zef2}. The magnetic properties of each sample were carefully
characterized to enable a comparison of their bulk properties. The inset of Fig. \ref{fig1}b shows the magnetization
as a function of temperature for a typical sample when field-cooled from above the LCMO ordering temperature (T${_C}$)
in 100Oe. The onset of ferromagnetism in LCMO is clearly observed at
$\sim$ 210K, along with the onset of superconductivity (T$_{SC}$), which for this particular sample (8 u.c. YBCO) was 80K. The samples
show similar properties to others  grown in the past, including the suppression of the superconducting
temperature with decreasing YBCO thickness (as shown in the inset of Fig. \ref{fig2}b), associated with the destruction of the superconducting pairing
mechanism due to the proximity to a ferromagnet \cite{pena1}. The ferromagnetic ordering of the Mn moment in the LCMO
layer is similar for all the samples, this is also shown in the inset of Fig. \ref{fig2}b.

XMCD measures the difference between X-ray absorption spectra for left- and right-handed
circularly polarized X-rays. To compensate for any beam instability, the XMCD measurements were performed using a double-difference technique.  The helicity of the light was reversed at 1Hz \cite{hara} whilst varying the incident 
energy under a particular fixed external magnetic field, immediately followed by an identical measurement performed with the externally
applied magnetic field in the opposite direction. With the photon helicity and magnetization parallel ($I^+$) and
antiparallel ($I^-$), information can be obtained about both the electronic structure ($I^++I^-$) and the magnetic properties ($I^+-I^-$). 
To discriminate between the LCMO and YBCO layers, the incident energies were calibrated to probe the L$_2$ and L$_3$ 
edges of both Mn and Cu. All the samples fabricated were investigated at temperatures 
above and below T$_{SC}$ of the YBCO layer, and in addition, several samples were investigated
with a finer temperature scan.


A typical XMCD spectrum at the Cu edges is shown in Fig.\ref{fig1}b as a function of incident energy at 100K. There is clearly a dichroic signal 
present in this sample (8 u.c. of YBCO). Mn dichroism was also observed at this temperature which is antiferromagnetically coupled to the Cu moment\cite{sup}, in agreement with previous work \cite{chak2,werner,bern}. Note that the Mn XMCD signal is an order of magnitude stronger. An induced Cu moment has previously been observed and attributed to superexchange across the interface
once the LCMO layer orders \cite{chak2,werner,bern}, and is present at all temperatures below the ferromagnetic transition. 
It is difficult to give a  quantitive analysis of the magnetic moments, but following the magneto optical sum rules\cite{thole,carra} and the analysis in Ref. \cite{chak2}, we estimate the magnetic moments of Mn and Cu at 28 K to be $\approx 1\mu_{B}$ and $\approx 0.013\mu_{B}$ respectively for the sample with 8 u.c. of YBCO.  Intriguingly, we observe a finite Cu orbital magnetic moment of about 20\%  of its spin moment.  
Thicker layers display
startlingly different magnetic properties. Fig. \ref{fig2}a shows the hysteresis loops for two different thicknesses (8 and 25 u.c.) of YBCO, measured with an incident X-ray energy of 931eV at 100K.  It can be seen clearly that the Cu moment observed at 100K in the YBCO normal state for the 8 u.c. sample is not present for the 25 u.c. sample.

Measurements of both samples as a function of temperature were performed to investigate the difference in the 
observed Cu moment. Fig. \ref{fig2}b shows the temperature dependence of the Cu and Mn moments in a sample with a
YBCO thickness of 8 u.c. and the Cu moment of the 25 u.c. YBCO sample, the data being obtained from XMCD measurements performed in a static field. The moment ($M$) in all cases is 
normalized to the lowest temperature measured ($M(0)$). It is clear that the Cu moment of the 8 u.c. sample tracks the Mn
moment. In contrast, the behavior of the Cu moment in the sample with 25 u.c. of YBCO is fundamentally different. Again, the onset of the Mn moment is observed 
below 200K (a behavior consistent across all the samples), but no Cu moment is observed in the normal phase of the YBCO. A 
Cu moment is only observed once the sample enters the superconducting region and thin films of only YBCO showed no evidence of any Cu moment above or below T$_{SC}$  

Further measurements on samples with varying  YBCO thicknesses were performed to investigate the normal
state of YBCO. Measurements were performed above T$_{SC}$ for all the samples at
150K and 100K respectively. Every sample exhibited a Cu
moment in the YBCO superconducting regime, and a Mn moment below the LCMO ordering temperature (see inset of Fig. \ref{fig2}b). 
However, the observation of a Cu moment in the normal phase of YBCO was found to depend on the YBCO layer thickness. We find that above
T$_{SC}$ only samples with a thickness of fewer than 12 u.c. show evidence of a Cu moment
(with seven investigated in this thickness regime), while six thicker samples (with YBCO thicknesses between 15 and 26 u.c.) showed no moment. 
Although there is an uncertainty here of three unit cells,
there appears to be a cut-off at which the superexchange interactions are suppressed in samples with an YBCO layer thickness greater than 12 unit cells.
Note that the observed thickness is not related to the suppression of T$_{SC}$ (see the inset of Fig.\ref{fig2}b). The Cu moment is expected to scale with the Mn moment for all samples and any Cu signal induced by the LCMO layer would be above the noise level of the XMCD signal irrespective of the YBCO layer thickness.
Furthermore, no effect on either the Cu or Mn 
moment is seen at 105K for all samples, the temperature at which the SrTiO$_3$ (substrate) is known to undergo a structural distortion.

To confirm the observation that the Cu moment is suppressed at T$_{SC}$, the 25 u.c. sample was annealed in an Ar atmosphere at 523K for 6 hours. This had the effect of reducing T$_{SC}$ to 64K without changing the ordering temperature of the LCMO, as confirmed by magnetometry measurements\cite{sup}. XMCD measurements were performed on this heat-treated sample, above and below the new T$_{SC}$. It is assumed that T$_{SC}$ has been lowered because of oxygen depletion.  This had the effect of 
damaging the surface of the sample, and indeed XMCD spectra at either the Cu or Mn edge accumulated using the TEY method could only be obtained after the removal of approximately 1 u.c. of LCMO by Ar sputtering in an adjacent experimental chamber. The pre-annealed sample showed a Cu moment at 70K, whilst the post-annealed sample exhibits no Cu moment at this temperature. 
Below 64K the Cu moment returned, indicating that the observation of a Cu moment appears to be connected with the 
superconductivity.

We have clearly demonstrated that the presence of the Cu moment in the YBCO normal state is dependent on its layer thickness. We confirm the existence of an induced moment in the interfacial Cu plane which, at least for an YBCO thickness below 12 u.c. is in agreement with other work \cite{chak2,werner}.  For  thicker samples the induced Cu moment only appears below the superconducting transition, suggesting the observed behavior is not a property of sample roughness or interdiffusion. We can rule out roughness as a dominant effect because when we examine a 30 u.c. YBCO bilayer the Cu moment in the normal state returns. Further work is required to 
understand any contribution from the roughness but this property alone cannot account for the observed behavior in our samples.
Our data appear not to be consistent with the work of Werner {\it et al.} \cite{werner}, but it should be noted that they examined a bilayer of different layer thicknesses which was grown via pulsed laser deposition rather than by DC sputtering. 

Below T$_{SC}$ there are several mechanisms that could result in a Cu moment. Magnetic domains can develop below the superconducting transition and have been attributed to the interaction between the layers being modified due to the presence of vortices creating electrodynamic forces \cite{chak2}. Another possibility is the presence of a triplet supercurrent at the interface creating an observable Cu moment \cite{echrig}, possibly due to an inhomogeneity of the surface magnetization. This prediction has recently been experimentally supported in YBCO/LCMO layers \cite{hu,dybko,kalcheim,visani}. 

These observations do not account for the lack of induced Cu moment in the high temperature normal phase of the thick YBCO films. A situation can be envisaged where a moment induced via the circulating current \cite{fauque} proposed by Varma  {\it et al.}  \cite{varma1, varma2} has its own exchange, independent of the interfacial exchange. Indeed orbital currents have recently been observed in CuO$_2$ plaquettes, the basic building block of cuprate superconductors \cite{science}. Our observations seem to suggest that in the normal phase of YBCO there is a characteristic thickness above which magnetic correlations can overcome the 
exchange energy arising from the LCMO layer. Indeed the cut-off regime is coincident with the known spin diffusion length in 
YBCO thin films which have been experimentally determined to be in the range of $8-11$ u.c.  \cite{nemes,soltan}. A change in the surface termination could change the interaction between the FM/SC bilayer.  However, this would still strongly indicate that there is an intrinsic magnetic coupling in the YBCO thin film at high temperatures which is dependent on the normal state of YBCO. Any connection to the psuedogap could be examined in the future by using films with a higher $T_{SC}$, as this would guarantee that the LCMO ordering temperature is above the psuedogap onset temperature thus testing if this phase can be used to explain the origin of the Cu moment suppression above $T_{SC}$.


In summary, we have performed XMCD measurements on bilayers of YBCO/LCMO, finding that above the superconducting temperature the observation of a Cu moment is dependent upon the thickness of the superconducting 
layer. The superexchange mechanism between the Mn and Cu in adjacent layers at the interface is not strong enough to compete with an apparent coupling mechanism in the YBCO layers which becomes dominant for layer thicknesses above 12 
u.c.. This direct measurement suggests that in the normal phase of YBCO magnetic fluctuations occur with a correlation length  of approximately 12 u.c. (14nm).



\section*{Acknowledgements}
Our experiments were performed with the approval of the Japan
Synchrotron Radiation Research Institute (JASRI) under Proposal Nos. 2007B1516 and 2008B1607). We thank K. Kodama for experimental assistance. We acknowledge financial support from the EPSRC (UK) under grant EP/G056463/1.  Work at UCM  supported by Spanish MICINN Grants MAT 2008 06517 and CSD2009-00013 (Imagine) and CAM S2009/MAT-1756 (Phama).


\begin{thebibliography}{99}
\bibitem{bg}
Blank samples were run to check the XAS peak position for pure Cu and Cu oxide.
\end{thebibliography}
\end{document}